\documentclass[sigconf]{acmart}
\copyrightyear{2020}
\acmYear{2020}
\setcopyright{acmcopyright}
\acmConference[CIKM '20] {The 29th ACM International Conference on Information and Knowledge Management}{October 19--23, 2020}{Virtual Event, Ireland}
\acmBooktitle{The 29th ACM International Conference on Information and Knowledge Management (CIKM '20), October 19--23, 2020, Virtual Event, Ireland}
\acmPrice{15.00}
\acmDOI{10.1145/3340531.3412050}
\acmISBN{978-1-4503-6859-9/20/10}
\usepackage{caption}
\usepackage{mathtools}
\usepackage{tikz}
\usepackage{multirow}
\usepackage{enumitem}
\usepackage{array}
\usepackage{color, colortbl}
\definecolor{Gray}{gray}{0.95}
\usepackage{tikz-dependency}
\usepackage{float}
\DeclareMathOperator*{\argmax}{arg\,max}
\newcommand{\matr}[1]{\mathbf{#1}}
\newcommand{\quotes}[1]{``#1''}
\DeclarePairedDelimiter{\norm}{\lVert}{\rVert}

\DeclarePairedDelimiterX{\infdivx}[2]{(}{)}{%
  #1\;\delimsize\|\;#2%
}

\newcommand*{\SuperScriptSameStyle}[1]{%
  \ensuremath{%
    \mathchoice
      {{}^{\displaystyle #1}}%
      {{}^{\textstyle #1}}%
      {{}^{\scriptstyle #1}}%
      {{}^{\scriptscriptstyle #1}}%
  }%
}

\newcommand*{\select}{\SuperScriptSameStyle{\bullet}}
\newcommand*{\qcm}{\SuperScriptSameStyle{\uparrow }}

\newcommand*{\ltrbase}{\SuperScriptSameStyle{\uparrow}}
\newcommand*{\ltrlda}{\SuperScriptSameStyle{\bullet}}

\begin{document}
\fancyhead{}

\title{Learning to Personalize for Web Search Sessions}

\author{Saad Aloteibi}
\authornote{This work was done while at the University of Cambridge.}
\affiliation{%
  \department{Department of Computer Science}
  \institution{King Saud University}
  \city{Riyadh}
  \country{Saudi Arabia}}
\email{SaadAloteibi@ksu.edu.sa}

\author{Stephen Clark}
\affiliation{%
  \department{School of Electronic Engineering and Computer Science}
  \institution{Queen Mary University of London}
  \city{London}
  \country{United Kingdom}}
\email{stephen.clark609@gmail.com}

\begin{abstract}
The task of session search focuses on using interaction data to improve relevance for the user's next query at the session level. In this paper, we formulate session search as a personalization task under the framework of learning to rank. Personalization approaches re-rank results to match a user model. Such user models are usually accumulated over time based on the user's browsing behaviour. We use a pre-computed and transparent set of user models based on concepts from the social science literature. Interaction data are used to map each session to these user models. Novel features are then estimated based on such models as well as sessions' interaction data. Extensive experiments on test collections from the TREC session track show statistically significant improvements over current session search algorithms. 
\end{abstract}

\maketitle

\section{Introduction}
\label{section:intro}
Web search is a dynamic process. Users interact with the results and reformulate their original query. Such an interaction can be considered as a form of feedback and it is a common behaviour \cite{pass2006}. Reformulation might occur for contrasting reasons. It could indicate a struggle in satisfying an information need or a success in locating relevant information for a specific aspect of a multi-facet information need and a move into researching another aspect. In either case, it is usually accompanied with behavioural actions that can signal latent variables about the user. \par

For example, one important user action happens when users click on a result and spend some time, known as \textit{click dwell time}, examining the clicked item. A study by Fox et al. \cite{fox2005} has found an association between click dwell time and user satisfaction. The longer the user stays on a clicked item the more likely that it satisfies her information need. Although these behavioural signals are noisy in nature \cite{joachims2007}, they present an opportunity to integrate unobtrusive users' behavioural information into various search engines' components. For instance, Agichtein et al. \cite{agichtein2006} have incorporated click-through and browsing features into ranking models and shown that it could provide significant improvement. In this paper, we follow a similar assumption for the task of session search. The goal of session search is to improve retrieval performance over a single search session. Each search session consists of multiple queries that are submitted by a user to fulfill a single information need. The assumption is that a user's interaction by means of reformulation and examination behaviour within the current session can be utilized to improve results for the session's next query. \par

Existing approaches to session search can be categorized into three classes. The first is to model the interaction process as a sequential decision making process where the goal is to learn a policy that maximizes a pre-defined reward which would lead eventually to improving the retrieval performance. One issue with such solutions is the choice of the reward function. They rely on using the ground truth judgments in their reward function. Typically, reward is defined in terms of nDCG$@$10 and the policy that maximizes the reward is used to rank documents. In practice, however, relevance assessment labels are not available to the agent. \par  

The second class of session search systems focuses on query formulation. These approaches use the reformulation sequence, click behaviour, previously presented documents and other available data to compose a new query or estimate a new query model that potentially better represents the user's information need compared with the session's current query \cite{guan2014, shen2005context, levine2017}. Such approaches, however, rely on the document's score against the newly formulated query as a single source of belief, or feature, about a document's relevance. It has become a standard for Web search engines to rank documents based on multiple features rather than fully relying on one scoring function or assuming a linear combination of features. Session search is no exception and many features can be extracted from the interaction information. The third class of systems approach session search as a learning to rank task. \par  

The main contribution of this paper is a personalization approach for session search under the learning to rank framework, which we refer to as LTR-SP. Given a set of pre-computed user models, each test session is mapped to its most relevant user models. These models provide richer context to define learning features and to identify related sessions from the query logs. Our approach extends beyond similar approaches. Previous work in this category uses features from the current session only \cite{jiang2014TREC, zhang2016ltr} or related sessions from the query logs submitted by the same user \cite{bennett2012} or other users \cite{li2015}. Our results show that our approach significantly outperforms existing session search approaches on four of session search's test collections. To the best of our knowledge, our approach is the first to provide consistent statistical improvement on all considered test collections. \par

\section{Related work}
\label{section:relatedWork}
Techniques from the reinforcement learning domain have been adopted in a number of previous session search studies. Previous work in this direction model session search using the framework of Markov Decision Processes (MDP) \cite{guan2013, chen2018} or its variant known as Partially Observable MDP (POMDP) \cite{luo2014win, luo2015DirectPolicy, yang2018}. The aim is to learn an optimal policy that maximizes the agent's reward. A policy prescribes action selection for the agent at each state. Guan et al. \cite{guan2013} propose the Query Change Model (QCM) as a session search retrieval model. QCM is based on MDP and treats queries as the system states. In this model, there are two agents: the user and the search engine. The user's actions are term retention, removal or addition while the search engine's actions are based on increasing, decreasing or maintaining query terms' weights. \par
 Luo et al. \cite{luo2014win} assume that the system states are hidden and thus model session search as a POMDP. Their proposed model, called Win-Win search, uses four hidden states based on two dimensions: relevance and exploration. The Win-Win system selects actions for the search engine from $20$ options that are based on various configurations of term weighting or retrieval models. Luo et al. \cite{luo2015DirectPolicy} also model session search using POMDP but learn optimal policies directly from a set of features that describes the observations that the search engine can make from the interaction process. Chen et al. \cite{chen2018} proposed a multi-agent MDP model where each agent is trained to rank documents for a specific cluster of related queries. The number of clusters and the number of MDP agents are determined according to a model based on the Chinese Restaurant Process framework. \par

Session search has been approached in a more standard IR way by treating interaction data as a source to expand or formulate the user's original query. Guan and Yang \cite{guan2014} explored various methods to aggregate all queries in a test session into a single query. A session's interaction data can be used to estimate a new query model. The query model $\theta_q$ is typically estimated using Maximum Likelihood Estimation (MLE) based on the current query only. In session search, the query chain and click data present additional contextual information that can be incorporated to estimate a new query model $\widehat{\theta_q}$. One popular approach to estimate $\widehat{\theta_q}$ is the Fixed Coefficient Interpolation (FixInt) method \cite{shen2005context}. It interpolates between two language models: the current query model $\theta_q$ and a history model $H$. The history model $H$ is also an interpolation between a click history $H_c$ and a query history $H_Q$. Levine et al. \cite{levine2017} constructs a query model $\widehat{\theta_q}$ by inductively interpolating a model of the current search iteration with its preceding queries. Both Shen et al. \cite{shen2005context} and Levine et al. \cite{levine2017} suggested methods to dynamically set interpolation parameters. \par

The third class of session search systems is based on the learning to rank framework. Bennett et al. \cite{bennett2012} consider three temporal views of a user's interaction. The first is a session view to capture a user's interactions within the current session. The second is a historic view that covers interactions prior to the current session and the third is an aggregate view. They defined a unified set of features that are calculated based on the three views. Their results suggest that the historic view provides significant improvement in personalizing the initial query of search sessions. As the session progresses, gain provided by the session view increases while the benefits of historic information decreases. Liu et al. \cite{liu2012} ran a laboratory study in order to identify a document's usefulness, or relevance, predictors. The most indicative predictor was found to be dwell time of clicked documents. They build decision tree models using dwell time and a few other variables to predict relevance. Documents that were judged as relevant were then used to extract expansion terms. Zhang et al. \cite{zhang2016ltr} cast the aggregation of the multiple contextual models extracted from consecutive interactions within a search session as a learning to rank problem. Several other studies have applied learning to rank algorithms to the task of session search \cite{ustinovskiy2013, shokouhi2013, jiang2014TREC}.   \par

\section{Learning to personalize for search sessions}
\label{section:approach}

\subsection{User models}
Our system approaches session search as a personalization task. An integral component of most personalization methods is the user model. We base our user models on role theory, one of the theoretical perspectives in the social science literature. In particular, we label each user model using its most relevant social position. A social position can be defined as \quotes{a collection of actors who are similar in social activity, ties, or interactions, with respect to actors in other position} \cite[p. 348]{wasserman1994}. The social position is treated as a community of users with similar interests. Examples of social positions include \textit{programmer}, \textit{traveller} and \textit{Red Sox fan}. For the purpose of this paper, we assume that a list of such social positions and a few seed terms for each social position is provided. We should note that such a list and seed terms are derived in an unsupervised manner from web documents. \par

We build a document collection for each social position by submitting the social position and its seed terms as a weighted query to our index of ClueWeb09. We further assume that the top $10$ documents for each position are pseudo-relevant. In order to learn a user model for each social position, we develop an extension to LDA \cite{blei2003}, called DiffLDA. The aim is to probabilistically model a social position from its document collection. We assume each social position's model is a multinomial distribution over words, i.e. a topic. Our model has the following three constraints: a social position's topic is highly likely to contribute in generating pseudo-relevant documents; topics different from the social position's topic may also be found in the pseudo-relevant set; and the social position's topic is not restricted to the set of pseudo-relevant documents but can contribute in generating the other documents. These are soft constraints and may be overridden during estimation if enough evidence is found in the data. We further assume the existence of a background topic that generates common terms in the social position document collection. Figure~(\ref{fig:gen_process}) presents the generative story of DiffLDA and Figure~(\ref{fig:difflda}) shows the model as a plate diagram.  \par

This model adds to the standard LDA model two components: the first is an observed binary variable $\Psi$ which is set to $1$ if the current document belongs to the pseudo-relevant documents. The second is a switching distribution $\Pi$ to facilitate capturing background terms. It is often the case that symmetric and low value priors are used when applying LDA to model documents, particularly to the $\alpha$ hyperparameter. The low value of such a parameter encourages the model to assign few topics for each document. In learning social positions' models, it would be preferable to bias the document topic distribution $\theta$ to include the position's topic as one of the topics for each document in the pseudo-relevance set. More formally, Let $R$ be the set of pseudo-relevant documents. Let $\boldsymbol{L}$ be a vector of length $K$ where $\boldsymbol{L}_m^k$ refers to the topic $k$ entry in vector $\boldsymbol{L}$ of document $m$. Similarly, $\boldsymbol{L}_m^r$ refers to the social position's topic, $r$, entry in vector $\boldsymbol{L}$ of document $m$. We set this vector as follows: $\boldsymbol{L}= \boldsymbol{0}$ if document $m \notin R$. $\boldsymbol{0}$ is a vector of all zeros of size $K$, effectively keeping the hyperparameter $\alpha$ unchanged. $\boldsymbol{L}= \boldsymbol{u}$ if document $m \in R$. $\boldsymbol{u}$  is a vector of all zeros of size $K$ except at $\boldsymbol{u}_r=\tau$, where $r$ is the index of the social position topic. The parameter $\tau$ act as a bias parameter that increases the probability of selecting the social position's topic for documents in the pseudo-relevant set. We set $\tau=2\alpha$. Both hyper-parameters $\alpha$ and $\beta$ are optimized using maximum likelihood estimation. The document topic hyperparamter $\alpha$ is then set according to step 8 (from Figure~\ref{fig:gen_process}) in the generative process for each document $m$. A similar transformation is used in the labeled LDA topic model \cite{ramage2009} where $\alpha$ is projected into a lower dimensional vector to restrict LDA to predefined topics. We perform parameter estimation using collapsed Gibbs sampling \cite{griffiths2004} for the joint probability in equation~\ref{eqn:collapsed}.
\begin{equation}
P(x,z,w) = P(x) \times P(z|x) \times P(w|z)
\label{eqn:collapsed}
\end{equation} \par

\begin{figure}
\centering
\caption{A graphical representation of DiffLDA.}
\label{fig:difflda}
\begin{tikzpicture}
\tikzstyle{main}=[circle, minimum size = 5mm, thick, draw =black!80, node distance = 6mm]
\tikzstyle{connect}=[-latex, thick]
\tikzstyle{box}=[rectangle, draw=black!100]

\node[main, fill=black!5] (alpha) {$\alpha$};
\node[main] (lambda) [above=of alpha] {$\lambda$};
\node[main] (pi) [below=of alpha] {$\Pi$};
\node[main] (mu) [left=of pi] {$\mu$};
\node[main] (theta) [right=of alpha] {$\theta$};
\node[main, fill=black!20] (psi) [above=of theta] {$\Psi$};
\node[main] (z) [right=of theta] {z};
\node[main, fill=black!20] (w) [right=of z] {w};
\node[main] (phi) [right=of w] {$\Phi$};
\node[main] (phib) [above=of phi] {$\Phi_b$};
\node[main] (phir) [below=of phi] {$\Phi_r$};
\node[main] (x) [below=of z] {$x$};
\node[main, fill=black!5] (beta) [right=of phi] {$\beta$};

 \path
 (lambda) edge [connect] (psi)
 (psi) edge [connect] (theta)
 (mu) edge [connect] (pi)
 (pi) edge [connect] (x)
 (x) edge [connect] (z) 
 (phib) edge [connect] (w)
 (phir) edge [connect] (w)
  (alpha) edge [connect] (theta)
  (theta) edge [connect] (z)
  (z) edge [connect] (w)
  (phi) edge [connect] (w)
  (beta) edge [connect] (phi);
  
\node[rectangle, inner sep=2.0mm,draw=black!100, fit= (z) (w) (x)] {};
\node[rectangle, inner sep=4.0mm,draw=black!100, fit= (z) (w) (theta) (x) (psi)] {};
\node[rectangle, inner sep=1.0mm,draw=black!100, fit= (phi)] {};

\node[rectangle, inner sep=1mm, fit= (z) (w) (theta) (x) (psi)  ,label=below left:\tiny{M}, xshift=0.5mm] {};  
\node[rectangle, inner sep=-1mm, fit= (z) (w) (x)  ,label=below left:\tiny{N}, xshift=0mm] {};  
\node[scale=1, rectangle, inner sep=-1.7mm, fit= (phi)  ,label=below right:\tiny{K}, xshift=-0.5mm] {};  

\end{tikzpicture}
\vspace{-4mm}
\end{figure}

\begin{figure}
\centering
\caption {The generative process of DiffLDA} 
\label{fig:gen_process} 
\begin{tabular}{p{5mm}p{70mm}}

1& Draw a switching prior distribution $\Pi \sim Dir(\mu)$ \\
2& For each topic $k \in \{1,...,k\}:$ \newline Draw a topic distribution $\Phi_k \sim Dir(\beta)$\\
3& Draw a background topic distribution $\Phi_b \sim Dir(\beta)$ \\
4& Draw a role topic distribution $\Phi_r \sim Dir(\beta)$ \\
5& For each document $m$:\\
6& For each topic $t \in \{1,...,T\}$\\
7& Draw $\Psi_m^t \sim Bernoulli(\lambda)$ \\
8& Generate $\boldsymbol{\alpha_m}= \boldsymbol{L_m} + \boldsymbol{\alpha}$ \\
9& Draw document topic distribution $\theta_m \sim Dir(\alpha_m)$ \\
10& For each word $n$ in document $m$:\\
11& Draw a switching variable $x_{m,n} \sim Multi(\Pi,1)$:\\
12& If $x_{m,n}=background$ : \newline Draw a word $w_{m,n} \sim Multi(\Phi_b,1)$ \\
13&If $x_{m,n} \neq background$ : \newline Draw a topic $\Phi_t \sim Dir(\beta)$ \newline Draw a word $w_{m,n} \sim Multi(\Phi_t,1)$ \\ 
\end{tabular}
\vspace{-4mm}
\end{figure}

\subsection{Initial ranker}
\label{section:initialRanker}

It is important to note that there are two search engines involved in our experiments. The first is a search engine used by the TREC session track organizers. This is the system with which users interacted. We refer to this system as \textit{the observer}. The second engine is the experimental system we use to index and retrieve from the document collections, referred to as \textit{local}. Both systems use different retrieval models and vary in their indexing parameters. Thus, results for each query also varies.\par

Learning to rank systems are usually applied to a sample of possibly relevant documents, i.e. re-ranking. In ad-hoc retrieval tasks, it is common to use standard retrieval models as the initial ranker. In session search, there are multiple queries for each session and richer contextual information about the user. It is sensible to apply a custom initial ranker that would increase the effectiveness of the initial results using a session's context. Therefore, we developed a simple initial ranker based on query formulation methods, which have been shown to be effective in session search \cite{guan2014, van2016}. Formally, the initial ranker query model $\widehat{\theta_q}$ is estimated as follows:
\begin{equation}
P(w | \widehat{\theta_q}) = \alpha P_{MLE}(w | \theta_{concat}) + (1 - \alpha) P(w | \phi_{q}^{*} )
\label{eqn:myInitRanker}
\end{equation}

$\theta_{concat}$ is a query model estimated over the concatenation of all queries in the session using maximum likelihood estimation. Van Gysel et al. \cite{van2016} found that concatenating all queries in a session led to improved performance. $\phi_{q}^{*}$ is a relevance model that is estimated based on one of the session's queries. Let $\phi_{i}^{observer}$ and $\phi_{i}^{local}$ represent the sets of relevance models for session $i$ by TREC search engine (observer) and local engine, respectively. We score each $\phi_{i,n}^{observer}$ based on the following function:
\begin{equation}
ModelScore(\phi_{i,n}^{observer}) =  \argmax_{\phi \in \phi_{i}^{local}} Jaccard(\phi_{i,n}^{observer}, \phi)
\label{eqn:scorePhi}
\end{equation}
Where $Jaccard(.,.)$ is the Jaccard coefficient between the two models' terms. $\phi_{q}^{*}$ is then selected as follows:
\begin{equation}
 \phi_{q}^{*} = \argmax_{\phi \in \phi_{i}^{observer} }  ModelScore(\phi)
\label{eqn:selectPhi}
\end{equation}
The goal is to select a set of expansion terms that both systems believe is relevant to the test session. The assumption here is motivated by Lee's \cite{lee1997} hypothesis in data fusion research. Lee argues that different retrieval models might return similar sets of relevant documents but not non-relevant documents. Similarly, if the two different systems produce a similar relevance model for a particular query in the test session, it is likely that this relevance model would be composed of expansion terms that are relevant to the sought-after session's information need and thus using such terms would likely improve the initial ranker performance. \par

\renewcommand{\arraystretch}{0.95}
\begin{table*}
\centering
\caption{Example sessions with their matched social positions and expansion terms. Tx.i refers to the $i^{th}$ session in TREC session track $x$. For presentation reasons, only the last two queries of each session are presented.}
\label{table:exampleMatchedSessions}
 \begin{tabular}{|p{0.05\linewidth}|p{0.24\linewidth}|p{0.15\linewidth}|p{0.46\linewidth}|}
	\hline
	\textbf{No.} & \textbf{Queries} & \textbf{Social positions} & \textbf{Social expansion terms} \\ \hline
T12.5 & pocono mountains park $\downarrow$ \newline  pocono mountains shopping & tour guide, tourist & accommodations, getaway, waterpark, attraction, honeymoon, camelback, vacation, trail, resort, hotel, trip, chalet, tour, lodging.\\ \hline
T14.44 & cyprus economic crisis $\downarrow$ \newline european economic crisis & economic consultant, policy analyst & eu, issues, eurozone, international, economy, cepr, imf, monetary, foreign, governance. \\ \hline
\end{tabular}
\end{table*}

\subsection{Matching search sessions to social positions}
\label{section:matching}

This section introduces a dynamic method to match each search session with a set of candidate social positions. The user model for each social position $p$ contains a static set of terms. It is unlikely that these term's comprehensively cover all aspects of the social position. Therefore, we rely on the concept of word embeddings to build a vector representation for each term in the test collection vocabulary. Having such a representation enables a similarity calculation between any term $t$ and social position $p$ by averaging over the similarity scores between $t$'s vector and the vectors of each term in $p$'s model. A document or a search session can be represented as a set of terms $T$. The task of matching a search session to a social position is then cast as finding the social position that is most similar to the terms set $T$. The assumption here is that terms that are relevant to social position $p$ should be semantically similar to $p$'s terms. An important by-product of this approach is that a set of social expansion terms that is relevant to the search session and to the matched social position can be jointly extracted. \par

Formally, let session $i$ be represented by a tuple $S_i = \langle Q_i, D_i, \phi_i, fb_i \rangle $. $Q_i=\{Q_{i,1}, Q_{i,2}, \dots, Q_{i,n}\}$  is the session's queries where $Q_{i,n}$ is the session's current query to be personalized. $D_i=\{D_{i,1}, D_{i,2}, \dots, D_{i,n}\}$ is the top $10$ documents returned for each query in the session. $\phi_i=\{\phi_{i,1}, \phi_{i,2}, \dots, \phi_{i,n}\}$ is the set of relevance models for the session. For instance, $\phi_{i,n}$ is computed over $D_{i,n}$ for query $Q_{i,n}$ in session $i$. $fb_i =\{fb_{i,1}, fb_{i,2}, \dots, fb_{i,n}\}$ is the set of expansion, feedback, terms for each iteration. It contains the top $n=40$ terms in the relevance model of each query. Note that there are different $D_i$ for the observer and local systems and thus different $\phi_i$ and $fb_i$. These relevance models are computed using RM1 \cite{Lavrenko2001}. \par

Let $T_i^{\text{observer}}$ denote the set of relevant terms for session $i$ using the observer system. $T_i^{\text{observer}}= \bigcup_{1 \leq j \leq n} fb_{i,j}$. A unified $T_i = T_i^{\text{observer}} \cup T_i^{\text{local}}$. In other words, $T_i$ contains expansion terms that the observer or the local systems believe are relevant to session $i$. Furthermore, let $R_j$ denote the set of terms representing the social position $j$. The similarity between social position $j$ and search session $i$ is then defined as follows:

\begin{equation}
Sim(T_i,R_j) = \frac{1}{|T_i|} \sum_{x \in T_i} \frac{1}{|R_j|} \sum_{z \in R_j} \frac{\matr{T_{i,x}} \matr{R_{j,z}}}{\norm{\matr{T_{i,x}}} \norm{\matr{R_{j,z}}} }
\label{eqn:sessionPosSim}
\end{equation}

where $T_{i,x}$ and $R_{j,z}$ are vector representations for terms $x \in T_i$ and $z \in R_j$. To obtain such a vector representation for each word, we use the continuous bag of words (CBOW) model \cite{mikolov2013}. The use of such vector representation means that equation~\ref{eqn:sessionPosSim} measures semantic similarity.\par

The second objective of this section is to identify a set of terms that are relevant to both the session and the session's social positions. Let $E_i$ denote this set such that $E_i \subseteq  T_i$. To populate $E_i$, we first calculate the probability that term $x \in T_i$ belongs to social position $j$ using the softmax function and equation~\ref{eqn:sessionPosSim}
\begin{equation}
P(j|x)= \frac{\text{exp} (Sim(x,R_j))}{\sum_{x' \in T_i} \text{exp} (Sim(x', R_j))}
\label{eqn:termPositionSoftMax}
\end{equation}
Let $j'$ be the most probable social position for $x$. $x$ will be added to $E_i$ if $j'$ is among the session's social positions. Table~\ref{table:exampleMatchedSessions} presents example sessions, their matched social positions and expansion terms.

\subsection{Identification of related search sessions}
\label{section:identification}

We assume that query logs might contain sessions that are similar to the current user's information need. The identification of such sessions is the focus of this section. Related sessions can be identified using term-based or content-based approaches. Luo et al. \cite{luo2014} constructed a term vector for each session from the combination of all queries in the session. Terms' idf values were assigned as weights. Vector representations for all sessions were then clustered using the k-means algorithm to discover topics in the query logs. Sessions belonging to the same cluster are considered related. Li et al. \cite{li2015} investigated the effectiveness of four classes of features with respect to: current query; query change; whole session; and related sessions. To determine topically related sessions, they estimate an LDA topic model over clicked documents in the entire query logs. They then build a topic vector for each session based on the session's clicked documents. The similarity between two sessions is calculated using the cosine similarity between their topic vectors. In essence, this task is an online task that needs to be performed at query time. The method proposed by Li et al. \cite{li2015} requires an LDA topic model to be estimated over all clicked documents which is prohibitive in practice. Also, these clustering methods require the number of topics to be set a priori. \par

We consider the task of identifying related sessions as a binary classification task with respect to a test session $t$. Formally, let $S$ be the set of all sessions in the query logs. $x_{e,t}$ is a feature vector to represent the relatedness, or lack of, between test session $t$ and $e \in S$. To train the classifier, we use the topic labels provided by the TREC session track organizers. If two sessions have the same topic label, they are considered related. We use the AROW algorithm to train the relatedness classifier \cite{crammer2009adaptive}. Table~\ref{tab:relatednessFeatures} presents the set of features. Pruning of unlikely candidate sessions is performed based on the following simple rule:
related sessions must have at least one social position in common and at least one shared result or query term. This pruning rule presupposes that for two sessions to be related, they must be relevant to at least one common social position. This captures the semantic similarity between two candidate sessions without the need to run a computationally expensive topic modelling algorithm on the entire clicked documents as in Li et al. \cite{li2015}. Features 8, 9 and 10 rely on identifying each session's social positions. The session's relevance model is the same relevance model used in the initial ranker and is estimated using equation~\ref{eqn:selectPhi}. The session's social relevance model is a vector of social expansion terms weighted using each term's weight in the session's relevance model. \par

\renewcommand{\arraystretch}{0.95}
\begin{table}
\vspace{-2mm}
\centering
\caption {Features used to identify related sessions. For features 4 and 5, a cutoff of $10$ results per query is applied.} 
\label{tab:relatednessFeatures}
\begin{tabular}{|p{\linewidth}|}
\hline
1. Number of shared query terms. \\
2. Ratio of shared query terms. \\
3. Number of identical queries in both sessions. \\
4. Number of shared results. \\
5. Ratio of shared results in both sessions. \\
6. Jensen-Shannon divergence between the two sessions' relevance model. \\
7. Jaccard similarity between the two sessions' expansion terms. \\
8. Jaccard similarity between the two sessions' social positions. \\
9. Jaccard similarity between the two sessions' social expansion terms. \\
10. Jensen-Shannon divergence between the two sessions' social relevance model. \\ \hline
\end{tabular}
\vspace{-2mm}
\end{table}

\subsection{Learning to rank features}
\label{section:features}

There are two sets of features. The first is features that are independent from the social positions of the test session while the second depends on identifying the session's social positions. Tables~\ref{tab:LTRFeatures} and~\ref{tab:SPLTRFeatures} list the features of both sets, respectively. In table~\ref{tab:LTRFeatures}, we consider five groups of features. The first four are query-dependent. These can be considered as different representations of the user's information need based on: the first and current queries; an aggregate query; the list of session's queries; and expansion terms. For each of those four groups, features are mostly based on scoring the relevance of the document, or the document's snippet, to the respective representation using the Query Likelihood model (QL) \cite{ponte1998}, BM25 \cite{robertson2009} and Hiemstra's language model (HLM) \cite{hiemstra1998}.  \par

\renewcommand{\arraystretch}{0.95}
\begin{table*}
\centering
\caption {Social position independent features.} 
\label{tab:LTRFeatures}
\begin{tabular}{|l|l|l|l|}
\hline
\textbf{Group} & \textbf{Feature} & \textbf{Description} &  \textbf{Total} \\ \cline{1-4} 
\multirow{2}{*}{First and Current}& FirstQuery/CurrentQuery        & Scores of first and current queries for document and snippet using scoring models.              &     12  \\ 
&AvgFirstAndCurrent & Average of first and current queries' scores using scoring models. & 3 \\ \cline{1-4} 
\multirow{8}{*}{Aggregate query}   &  AggregateQueryLength       &    Number of tokens in query.         &   1     \\ 
								& No.DistinctTerms    & Number of distinct terms.    &  1 \\ 
								& No.MatchedTerms    & Number of query terms in document.    & 1  \\ 
								& AggregateQueryRatio    &  Query's terms ratio in document and snippet.   &  2 \\ 
								& AggregateQueryScore    &  Query's scores for document and snippet using scoring models.   &  6 \\  												& TermStatistics   	 &  	Terms' statistics using scoring models.	  &  15 \\  		
								& TopTermsScores   	 &  Top terms' scores using scoring models. 		&  3 \\ 						
                             &    QueryModelScore     &     Query model's scores using scoring models.        &    3    \\ \hline
 \multirow{2}{*}{Session}   &  No.Queries       &   Number of queries in the session.         &    1    \\ 
							 &  SessionStatistics       &  Session's statistics using scoring models.          &   15     \\ \hline
 \multirow{3}{*}{Expansion}   &  ExpansionScores       &  Expansion terms' scores using scoring models.          &    3    \\ 
							 &  ClickedExpansionScores       &   Clicked documents' expansion terms using scoring models.         &   3     \\ 
							&  DocumentRank       &  Rank using the initial ranker as in equation~\ref{eqn:myInitRanker}.          &   1     \\ \hline
 \multirow{2}{*}{Document}   &  PageRank/Spamness       & PageRank \cite{page1999} and spamness \cite{cormack2011} scores.           &    2    \\ 
							 &  Stopwords/DocLength       &   Stopwords ratio and document length.         &   2     \\ 
							&  Wikipedia       &  Binary indicator for Wikipedia documents.          &   1     \\ \hline
\end{tabular}
\vspace{-2mm}
\end{table*}

\renewcommand{\arraystretch}{0.95}
\begin{table*}
\centering
\caption {Social position dependent features.} 
\label{tab:SPLTRFeatures}
\begin{tabular}{|l|l|p{0.52\linewidth}|l|}
\hline
\textbf{Group} & \textbf{Feature} & \textbf{Description} &  \textbf{Total} \\ \cline{1-4} 
\multirow{5}{*}{Related sessions}&TopicQueryTerms         &Number and ratio of topic query's terms in document.              &     2  \\ 
				& TopicQueryScores        & Topic query's scores using scoring models.             &   3    \\ 
				& TopicExpandTerms        &Ratio of topic's expansion terms in document.              & 1      \\ 
				&  TopicRelevanceScores       &Topic relevance model scores using scoring models.               &    3   \\ 
				&  ClickRelevanceScores       &Click relevance model scores using scoring models.             &   3    \\ \cline{1-4} 
\multirow{7}{*}{Social positions}&SocialExpandTerms         &   Number and ratio of social expansion terms in document.           &  2     \\ 
				& SocialExpandTermsTitles        &Scores of social expansion terms which appear in clicked documents' titles using scoring models.             &  3     \\ 
				& SocialExpandTermsSnippets        &Scores of social expansion terms which appear in clicked documents' snippets using scoring models.             & 3      \\ 
				& TopicSocialExpandTerms        &Ratio of the topic-level social expansion terms in document.              &     1  \\ 
				& SocialRelevanceScores        &Social relevance model scores using scoring models.              &  3     \\ 
				& TopicSocialRelevanceScores        &Scores of the topic-level social relevance model using scoring models.              &   3    \\ 
				& TopicSocialRelevanceScores[Clicked]        &Scores of the topic-level social relevance model, that is estimated over clicked documents, using scoring models.              &   3    \\ \hline
\end{tabular}
\vspace{-2mm}
\end{table*}

Our system focuses on using a session's interaction data to personalize the results of the user's current query, i.e. the last query in the session. It is thus intuitive to include features that represent the relevance of candidate documents to the current query. It is also logical to assume that some of the session's queries might capture the user's information need better than others. Guan and Yang \cite{guan2014} investigated the question of which queries in session search are more important and thus should be assigned higher weights in an aggregation scheme. Beside the current query, they found that the first query is almost as important as the current query. We include features to specifically account for the session's first query. \par

The second group is informed by previous research on query formulation, which focuses on composing a new query using the session's query chain. In particular, we include features representing two methods to build the new query. The first composes the new query as the concatenation of the reformulation chain. This is called an aggregate query. The second method estimates a query model by interpolating a query model built over the session history $H_Q$ with another query model $\theta_q$ that is built using the current query. The session history includes all queries prior to the current query. Formally, the query model is built using the following equation:
\begin{equation}
P(w | \widehat{\theta_q}) = \lambda P_{MLE}(w | \theta_q) +  (1-\lambda) P_{MLE}(w | H_Q) 
\label{eqn:LTRQueryModel}
\end{equation}

In addition, statistical relevance features for all the terms that the user used during the session are collected. These include: maximum, minimum, average, variance and standard deviation. The top terms features are meant to represent the most frequent term or terms in the query chain. These are terms that the user insists on including the most during the session.  The third group represents expansion terms, which are extracted using two approaches. The first is based on the relevance model selected using equation~\ref{eqn:selectPhi} as in the initial ranker component, section~\ref{section:initialRanker}. For the second approach, we estimate a relevance model using RM1 \cite{Lavrenko2001} over the session's clicked documents\footnote{All clicked documents are considered regardless of the dwell time.}. For both approaches, a cut-off of $40$ terms is applied. In addition, we include the document rank based on the initial ranker component. \par

The session features measure the relevance of each candidate document to each query in the session. This is approached by collected the following statistical values: maximum, minimum, average, variance and standard deviation. Finally, document features represent candidate document's quality features such as its PageRank score and spamness score. \par

In table~\ref{tab:SPLTRFeatures}, we consider two groups of social position dependent features: related sessions and social positions. In section~\ref{section:identification}, we discussed the design of a relatedness classifier to identify related sessions. The classifier identifies all related sessions to a test session. Relations between such identified related sessions and the test session are assumed to be transitive. Thus, the results of classifying all sessions in the query logs is a set of hard clusters. Each cluster is called a topic. The related sessions features measure the relevance of candidate documents to four novel representations: topic query; topic relevance model; topic expansion terms; and topic-clicked relevance model. The topic query is the concatenation of all related sessions' queries and the current session's queries. An inherent assumption is that the method by which related sessions are identified is fine-grained so that related sessions are about the same information need. Thus, the topic query would likely contain repeated terms that might represent the key term, or theme terms, of such an information need. The topic relevance model $\theta_t$ is computed as follows:

\begin{equation}
P(w|\theta_t)= \frac{\sum_{s \in R} P(w | \theta_s) }{|R|}
\label{eqn:topicRelevanceModel}
\end{equation}

where $R$ is the set of related sessions and $\theta_s$ is the relevance model for session $s$ as selected using equation~\ref{eqn:selectPhi}. Topic expansion terms are the top $40$ terms in the topic relevance model $\theta_t$. The topic-clicked relevance model is estimated using the RM1 model \cite{Lavrenko2001} over all clicked documents in sessions that belong to the same topic cluster as the current session.  \par

In section~\ref{section:matching}, we described a method to match the current session $i$ to its most relevant social positions. One objective of this matching process was to identify a set of terms $E_i$ that is likely to be relevant to the current session and its social positions. We call this set the social expansion terms and examples of them are shown in table~\ref{table:exampleMatchedSessions}. A further three sets of social expansion terms are derived and used to introduce features in table~\ref{tab:SPLTRFeatures} . The first is a subset of $E_i$ containing terms that occur in the current session clicked documents' titles and another one for terms appearing in the snippets of clicked documents. The assumption is that these two sets would contain highly relevant terms to both the session and its social positions that could have triggered the user to click on these documents. The third set contains topic-level social expansion terms which is built as the union of all the social expansion terms for sessions that belong to the same topic $R$ as $\bigcup_{s \in R} E_s$. \par

The social position features also include three types of relevance models. The first is the social relevance model. For session $i$, this model's terms are the social expansion terms $E_i$. Terms are weighted based on the session's relevance model. The session relevance model, discussed in section~\ref{section:initialRanker}, is the RM1 model \cite{Lavrenko2001} for one of the session's queries as selected by equation~\ref{eqn:selectPhi}. The weighted average of all social relevance models for sessions that are members of the same topic forms a topic-level social relevance model. It is constructed in a similar way as the topic relevance model in equation~\ref{eqn:topicRelevanceModel} except that its components must be in $E_i \: \forall \: i \in R$. Finally, a third variation is estimated using clicked documents in sessions that are related to the current session. The components of this model are also limited to social expansion terms only. \par

\section{Experimental setup}
\label{section:setup}
In this paper, we aim to answer the following research questions: \textbf{(RQ1)} How effective is the proposed learning to rank approach for session search compared with other well-established systems? \textbf{(RQ2)} What is the significance of features that are estimated using social positions' models? \textbf{(RQ3)} Does the identification and use of related sessions' data improve performance? And how effective is the proposed relatedness classifier in identifying related sessions compared with an alternative approach? \textbf{(RQ4)} Which sessions are better personalized than others using the proposed approach?\par
We evaluate the proposed approach on TREC 2011-2014 session tracks \cite{carterette2016}. The TREC session tracks provide three types of contextual information. The first is the sequence of queries leading up to the test query for each session. The second is the ranked list of documents for each past query. The third is the user's clicking behaviour. The relevance of a document was judged based on the topic description, i.e. whole session relevance. There are $1,282$ sessions in total. We use these sessions for the identification of related sessions in section~\ref{section:identification}. The session tracks of TREC11 and TREC12 used ClueWeb09 as their document collection and ClueWeb12 for the 2013 and 2014 tracks. We use category B of both collections in our experiments, which is 50 million pages per collection. In addition, the organizers of the TREC14 session track provided a baseline run to each participant to use. We indexed all the documents that are included in the baseline run since some of them are not included in category B of ClueWeb12. All experiments on TREC14 are based on the organizers' baseline run. We use a custom built retrieval system and stem queries and documents using the Krovetz stemmer \cite{Krovetz1993}. Documents with a spam percentile less than $70$ are removed \cite{cormack2011}. This list of candidate documents is truncated at rank $100$ for all queries and is then used by the learning to rank model to produce the final runs. The evaluation metrics used in this paper are based on TREC session track's official metrics. These are: nDCG@k, nERR@k  and MAP. All runs are evaluated using the official evaluation script. Statistical tests are performed using paired t-test ($p < 0.05$). \par

To validate our approach, we compare with the following systems in addition to the initial ranker:
\begin{itemize}
\item \textbf{Current query:} A retrieval system based on the QL model \cite{ponte1998} with a Dirichlet smoothing parameter $\mu = 3500$. This system uses the current query only.
\item \textbf{Best TREC:} This baseline refers to the best performing runs for each TREC session track.
\item \textbf{Aggregated query:} A concatenation of all the session's queries as suggested in Van Gysel et al. \cite{van2016}. The retrieval model is QL with similar settings as in the current query baseline.
\item \textbf{QCM:} The Query Change Model as used by Guan et al. \cite{guan2013}. 
\end{itemize}

In terms of parameters, a trade-off parameter $\alpha$ is used to interpolate between the concatenated query model $\theta_{concat}$ and the best query's relevance model   $\phi_q^*$ in the initial ranker equation~\ref{eqn:myInitRanker}. A similar parameter $\lambda$ is used to interpolate the current query model $\theta_q$ with a history query model $H_q$ in equation~\ref{eqn:LTRQueryModel}. The resultant query model is used as a feature for the learning to rank model. Both parameters were set to an equal value ($\alpha =\lambda = 0.70$). We train our learning to rank model using the lambdaMART algorithm \cite{wu2008ranking}. We perform a 10-fold cross validation by splitting queries into training ($60\%$), validation ($20\%$), and test ($20\%$) sets. We use the lambdaMART implementation in the RankLib\footnote{\url{https://sourceforge.net/p/lemur/wiki/RankLib/}} library with default parameters. Statistical significance tests are performed in comparison to:  initial ranker (as an example of query formulation approaches) and QCM (reinforcement learning). \par

\section{Results and discussion}
\label{section:results}

\subsection{System validation}
\label{section:approachValidation}

\renewcommand{\arraystretch}{0.95}
\begin{table}
\centering
\caption{Search accuracy on session tracks.}
\label{table:SessionResults}
\begin{tabular}{llll}
\toprule
        & \textbf{nDCG@10}     &  \textbf{nERR@10}     & \textbf{MAP} \\ \midrule
        \multicolumn{4}{l}{\textbf{TREC 2011:}}   \\ \midrule
current			&	0.3480	&			0.3968	&			0.0824	\\ \midrule
Aggregated	&	0.4066 	(16.84\%)	 &	0.4644 (17.04\%)	&	0.1031 \\ \midrule
Initial&	0.4427\qcm 	(27.21\%) &	0.4980\qcm	(25.50\%)		&	0.1097 \\ \midrule
Best TREC \cite{best2011}		&	0.4540 (30.46\%)	&	0.5208\qcm	(31.25\%)	&	0.1253 \\ \midrule
QCM				&	0.4079 (17.21\%)		&	0.4550  	(14.67\%)	&	0.1130\select	\\ \midrule
LTR-SP			&	\textbf{0.5006}\qcm\select	(43.85\%)		&	\textbf{0.5698}\qcm\select 	(43.60\%)	&	\textbf{0.1335}\qcm	\\ \bottomrule
\multicolumn{4}{l}{\textbf{TREC 2012:}}   \\ \midrule
current			&	0.2478	&			0.2991	&				0.1183	\\ \midrule
Aggregated	&	0.2941	(18.68\%)		&	0.3449	(15.31\%)		&	0.1387	\\ \midrule
Initial&	0.3464\qcm	(39.79\%)		&	0.3899\qcm	(30.36\%)	 &	0.1576\qcm 	\\ \midrule
Best TREC	 \cite{jiang2012TREC}	&	0.3221 (29.98\%) 	&	0.3595	(20.19\%)		&	0.1457\qcm	\\ \midrule
QCM				&	0.2746 (10.82\%)		&	0.3218	(7.59\%)	&	0.1169	\\ \midrule
LTR-SP			&	\textbf{0.3907}\qcm\select	(57.67\%)	&	\textbf{0.4748}\qcm\select 	(58.74\%)	&	\textbf{0.1620}\qcm	\\ \bottomrule
        \multicolumn{4}{l}{\textbf{TREC 2013:}}   \\ \midrule
current			&	0.1000	&		0.1337	&		0.0322	\\ \midrule
Aggregated	&	0.1302	(30.20\%)	&		0.2031	(51.91\%)	&		0.0443	\\ \midrule
Initial &	0.1303	(30.30\%)	&		0.2010	(50.34\%)	&		0.0448	\\ \midrule
Best TREC	 \cite{jiang2013TREC}	&	0.1706\select	(70.60\%)	&		0.2049	(53.25\%)	&	\textbf{0.0873}\qcm\select	\\ \midrule
QCM				&	0.1480	(48.00\%)	&	0.2108	(57.67\%)	&		0.0436	\\ \midrule
LTR-SP			&	\textbf{0.1893}\qcm\select	(89.30\%)	&	\textbf{0.2817}\qcm\select	(110.70\%)	&		0.0593\qcm\select	\\ \bottomrule
        \multicolumn{4}{l}{\textbf{TREC 2014:}}   \\ \midrule
current			&	0.1937	&			0.2263	&			0.0819\qcm	\\ \midrule
Aggregated	&	0.2028	(4.70\%)	&		0.2489	(9.99\%)	&		0.0827\qcm	\\ \midrule
Initial&	0.2125	(9.71\%)	&	0.2596	(14.71\%)	&		0.0846\qcm	\\ \midrule
Best TREC \cite{luo2014}		&	0.2580	(33.20\%)	&		0.3268\select	(44.41\%)	&		0.0730	\\ \midrule
QCM				&	0.2443	(26.12\%)	&		0.3126	(38.14\%)	&		0.0636	\\ \midrule
LTR-SP			&	\textbf{0.3222}\qcm\select	(66.34\%)	&		\textbf{0.4145}\qcm\select	(83.16\%)	&	\textbf{0.1055}\qcm	\\ \bottomrule
\end{tabular}
\vspace{-2mm}
\end{table}

In this section, we address the first research question (RQ1). To validate the effectiveness of our proposed approach LTR-SP, we compare its performance to other related approaches, including state-of-the-art systems. Results using the TREC 2011, 2012, 2013 and 2014 test collections are presented in table~\ref{table:SessionResults}. Statistical significant improvement over the initial ranker and the QCM system are denoted with the symbols (\select) and (\qcm), respectively. The change percentages compared with the current query model are reported for all runs. The best scores are highlighted in bold for each test collection.  \par

LTR-SP performs substantially better than all other systems in terms of nDCG@10 and nERR@10. LTR-SP also provides the best MAP scores for all the test collections except for TREC 2013. LTR-SP improvements over the QCM and the initial ranker baselines are statistically significant with regard to nDCG@10 and nERR@10 on all datasets. Furthermore, LTR-SP significantly outperforms QCM on MAP as well for all test collections. The change percentages across all years also provide further support for LTR-SP. For instance, LTR-SP improvements relative to the best TREC system are $10.26\%$, $21.30\%$, $10.96\%$ and $24.88\%$ in terms of nDCG@10 for TREC 2011, 2012, 2013 and 2014, respectively. Under nERR@10, LTR-SP improvements over the best TREC system are by $9.41\%$, $32.07\%$, $37.48\%$ and $26.84\%$ on TREC 2011 to 2014. This trend also holds when comparing LTR-SP with the other approaches for both metrics nDCG@10 and nERR@10 and for MAP except on TREC 2013. For TREC 2013, most relevant documents come from ClueWeb12 category A rather than its subset category B that is used by LTR-SP. This, perhaps, explains the difference in terms of MAP between the best TREC system that uses category A and LTR-SP. \par

Besides the average gain that LTR-SP provides under the considered evaluation metrics, it is important to analyze the robustness of LTR-SP based on the amount of queries that are positively and negatively affected by this approach. Again, LTR-SP excels over the other approaches. If we consider all the $361$ test queries, LTR-SP improves performance on about $64\%$ of them. This is the highest percentage of positively affected queries. It is followed by the best TREC system at $54\%$ and QCM at $50\%$. In terms of hurt queries, LTR-SP has the lowest percentage at $19\%$ compared with the best TREC system at $28\%$, the initial ranker at $28\%$ and the QCM at $30\%$. Overall, the consistency of LTR-SP strong performance on these four test collections and its statistically significant improvement over the baselines under nDCG@10 and nERR@10 provide sufficient evidence that LTR-SP is an effective session search system. \par

\subsection{Social positions' models}
\label{section:modelsResults}

With regards to our second and third research questions, we perform an ablation study to investigate the significance of features that are estimated using social positions' models and the use of an alternative method to identify related sessions. LTR-Base is a learning to rank model that is trained using general features which do not depend on inferring the session's social position. These features are listed in table~\ref{tab:LTRFeatures}. LTR-LDA uses the related sessions' features in table~\ref{tab:SPLTRFeatures} in addition to all the general features in table~\ref{tab:LTRFeatures}. Related sessions were identified using the LDA-based approach that was discussed in section~\ref{section:identification} \cite{li2015}. These systems are compared with LTR-SP which uses the full set of features. The performance of these systems is shown in table~\ref{table:ltrSystems}. As can be noted from this table, LTR-SP significantly outperforms both LTR-Base and LTR-LDA under the three evaluation metrics. There are three main observations from this table. Firstly, the performance of LTR-Base is better than all the other baselines. This, perhaps, indicates the usefulness of the general features in table~\ref{tab:LTRFeatures}. It also suggests that a learning to rank approach could be a viable solution for session search. Secondly, this table shows an improvement of about $3.55\%$ on nDCG@10 as a result of including related sessions' features. Similar trends are also observed in terms of nERR@10 and MAP. These improvements support the claim that related sessions' features are useful. Thirdly, the performance of LTR-SP is still superior to that of the LTR-LDA. LTR-SP achieves $0.3463$ on nDCG@10, $0.4316$ on nERR@10 and $0.1156$ on MAP. These are higher than LTR-LDA by $5.90\%$, $7.15\%$ and $5.19\%$, respectively. While LTR-SP has additional social position specific features, it identifies related sessions more efficiently than LTR-LDA. The latter requires estimating an LDA model for all clicked documents in the query logs. This is an expensive operation to be performed at query time. In LTR-SP, the identification of related sessions was formulated as a binary classification task with a few informative features that can be done online with minimal overhead.  \par 

\renewcommand{\arraystretch}{0.95}
\begin{table}
\centering
\caption{Performance of learning to rank approaches on 361 test sessions from TREC session tracks 2011-2014. Significant improvement over LTR-Base is denoted with (\ltrbase) while (\ltrlda) indicates significant improvement over LTR-LDA.}
\label{table:ltrSystems}
\begin{tabular}{lccc}
\toprule
        & \textbf{nDCG@10}  & \textbf{nERR@10} & \textbf{MAP}  \\ \cline{1-4}
         QCM		&	0.2638 	& 	0.3205 	& 	0.0836		\\ 
          Best TREC		&	0.2956	& 	0.3471	& 	0.1072		\\ 
 LTR-Base		&	0.3158	& 	0.3861	& 	0.1073 		\\ 
 LTR-LDA		&	0.3270\ltrbase	& 	0.4028\ltrbase	& 	0.1099		\\ 
 LTR-SP			&	\textbf{0.3463}\ltrbase \ltrlda	& 	\textbf{0.4316}\ltrbase \ltrlda	& 	\textbf{0.1156}\ltrbase \ltrlda 		\\ \cline{1-4}
\end{tabular}
\vspace{-2mm}
\end{table}

To further understand the relative contribution of features that are estimated using social positions' models, we calculate the Gini impurity \cite{giniIndex} for all trees averaged over all 10 cross-validation splits. These scores are then normalized relative to the feature with the highest Gini index. Table~\ref{table:GiniAnalysis} lists the top $25$ features. Features that use social positions' models are highlighted in bold. About $48\%$ of these top features are social position dependent and $52\%$ are general features. An examination of this list reveals the important role of social positions' features. First, about $40\%$ of all social positions' features are placed at the top 25 features compared with only $17\%$ general ones. Secondly, there are $12$ types of social positions' features as shown in table~\ref{tab:SPLTRFeatures}. $9$ out of those $12$ have at least one feature in table~\ref{table:GiniAnalysis}. This analysis and the ablation study demonstrates the significance of social positions' features and answers the second research question (RQ2). It also provides evidence for the usefulness of related sessions' features (RQ3). \par

\renewcommand{\arraystretch}{0.95}
\begin{table}
\centering
\caption{The relative importance of features based on their Gini index scores. }
\label{table:GiniAnalysis}
\begin{tabular}{lc}
\toprule
 \textbf{Feature} & \textbf{Gini index} \\ \toprule
1. Stopwords.																				 				 &			$1.000$										 \\ 
2. DocumentRank.					 &			$0.430$									 \\ 
\textbf{3. TopicQueryScores[BM25].} 												 		 &			$0.301$								 \\ 
4. Spamness.												 &			$0.296$										 \\ 
\textbf{5. TopicRelevanceScores[BM25].}											 	 &			$0.227$									 \\ 
\textbf{6. ClickRelevanceScores[BM25].}										 &			$0.226$									 \\ 
\textbf{7. TopicQueryScores[QL].}																 &			$0.154$									 \\ 
8. AggregateQueryScore[Snippet][BM25].		 																		 &			$0.144$	\\ 
\textbf{9. SocialExpandTermsSnippets[BM25].}											 &		$0.135$ \\ 
10. TermStatistics[Maximum][BM25].									 &					$0.123$								 \\ 
11. ExpansionScores[BM25].										 &						$0.104$										 \\ 
12. FirstQuery[Snippet][BM25].										 &				$0.080$				 \\ 
13. CurrentQuery	[Snippet][BM25].											 &				$0.069$			 \\ 
\textbf{14. SocialExpandTermsTitles[BM25].}			 		 &							$0.068$						 \\ 
\textbf{15. TopicSocialRelevanceScores[Clicked][HLM].}														 &							$0.041$		 \\ 
16. SessionStatistics[Maximum][BM25]. 									 &		$0.040$											 \\ 
17. TermStatistics[Average][BM25].							 &		$0.038$											 \\ 
\textbf{18. TopicExpandTerms.}							 &				$0.035$									 \\ 
\textbf{19. SocialExpandTermsTitles[HLM].}	 &		$0.034$											 \\ 
20. AvgFirstAndCurrent[QL].											 &			$0.033$										 \\ 
21. FirstQuery[Document][QL].					 &								$0.031$					 \\ 
\textbf{22. TopicSocialExpandTerms.}								 &			$0.028$										 \\ 
\textbf{23. TopicSocialRelevanceScores[BM25].}									 &		$0.028$											 \\ 
\textbf{24. TopicSocialRelevanceScores[Clicked][QL].}	 &		$0.026$											 \\ 
25. SessionStatistics[Maximum][QL].										 												 &	 $0.024$												 \\ \bottomrule
\end{tabular}
\vspace{-2mm}
\end{table}

The significance of social positions' features is twofold.  First, social positions' models play a central role in efficiently and effectively identifying related sessions that are likely to be issued by a user with a similar information need. A session's social positions provide features for the classifier that identifies related sessions, as explained in section~\ref{section:identification},  as well as a rule for the early pruning strategy. The third and seventh important features in table~\ref{table:GiniAnalysis} are about the topic query. As mentioned earlier, the topic query is a concatenation of all related sessions' queries. Concatenating unrelated queries would likely harm rather than improve the result's relevance. Thus, it is critical to identifying only closely related sessions. Second, in section~\ref{section:matching}, we introduced a method to extract social expansion terms while ensuring their relevance to both the session and its social positions. These terms proved to be useful especially if they appear in the titles or snippets of clicked documents as in features 9, 14, 18, 19 and 22 in table~\ref{table:GiniAnalysis}. In addition, we use social expansion terms in building a social relevance model for the session and a variant of such a model that is estimated using social relevance models for all related sessions. Three out of the top 25 features are based on the topic-level social relevance model.  \par

The performance of the social positions based classifier and the LDA-based approach can be measured using the evaluation metric F1 based on the gold standard mapping. The social positions' classifier achieves an F1 score of $0.77$ compared with a $0.30$ for the LDA-based classifier.  This low F1 score is expected. The LDA-based approach only used clicked documents to estimate an LDA model. Few sessions have clicked documents and the number of clicked documents per session is typically small. \par

\subsection{Types of search sessions}
\label{section:typesResults} 
Next, we answer the fourth research question (RQ4) in regard to analyzing the performance of LTR-SP and other approaches on session types. Starting from TREC 2012, sessions were classified based on two facets using a framework introduced by Li and Belkin \cite{li2008}. These two facets are: product and goal. The product of a session can be merely locating facts or information items on the Web. This is called a factual product. It can also be an intellectual product when it results in new ideas or findings. The goal of the session can be either specific or amorphous. These two facets produce four types of sessions: known-item (factual specific), interpretive (intellectual specific), known-subject (factual amorphous) and exploratory (intellectual amorphous). Across TREC 2012, 2013 and 2014 test collections, there are 112 known-item ($39.30\%$), 56 interpretive ($19.65\%$), 59 known-subject ($20.70\%$) and 58 exploratory sessions ($20.35\%$). \par

\renewcommand{\arraystretch}{0.95}
\begin{table}
\centering
\caption{Performance on four sessions types \cite{li2008}.}
\label{table:resultsBySessionsType}
\begin{tabular}{llll}
\toprule
        & \textbf{nDCG@10}  & \textbf{nERR@10} & \textbf{MAP}  \\ \cline{1-4}
\multicolumn{4}{l}{\textbf{Factual specific (known-item)}}\\ \cline{1-4}
 TREC		&	0.2550  &	0.2798	 &	\textbf{0.1159} 		\\ 
 QCM				&	0.2132	 &	0.2621	 &	0.0786 		\\ 
 LTR-Base		&	0.2659	&	0.3332	 &	0.1058 	\\ 
 LTR-LDA		&	0.2798	 &	0.3596	&	0.1072 		\\ 
 LTR-SP			&	\textbf{0.2932}  &	\textbf{0.3780}	 &	0.1111 	\\ \cline{1-4}
\multicolumn{4}{l}{\textbf{Factual amorphous (known-subject)}} \\ \cline{1-4}
 TREC		&	0.2720	&	0.3437  &	0.1233 	\\ 
 QCM				&	0.2215	 &	0.2850	 &	0.0920 	\\ 
 LTR-Base		&	0.2839	 &	0.3822	 &	0.1285 		\\ 
 LTR-LDA		&	0.3119	&	0.4036 &	0.1336 	\\ 
 LTR-SP			&	\textbf{0.3203}	&	\textbf{0.4196}	 &	\textbf{0.1462} 		\\ \cline{1-4}
\multicolumn{4}{l}{\textbf{Intellectual specific  (interpretive)}} \\ \cline{1-4}
 TREC		&	0.2194	 &	0.2849	 &	0.0522  		\\ 
 QCM				&	0.1791 &	0.2396	 &	0.0358 	\\ 
 LTR-Base		&	0.2474	&	0.3210	 &	0.0629 		\\ 
 LTR-LDA		&	0.2514	&	0.3329	&	0.0645 		\\
 LTR-SP			&	\textbf{0.2811}&	\textbf{0.3723}	 &	\textbf{0.0716} \\ \cline{1-4}
\multicolumn{4}{l}{\textbf{Intellectual amorphous (exploratory)}} \\ \cline{1-4}
 TREC		&	0.2640	 &	0.3130	 &	0.1033 \\ 
 QCM				&	0.2974	&	0.3715	 &	0.0926 	\\ 
 LTR-Base		&	0.3163	 &	0.3805	&	0.1035 \\ 
 LTR-LDA		&	0.3265	&	0.3951	&	0.1078 	\\
 LTR-SP			&	\textbf{0.3362}	&	\textbf{0.4234}	&	\textbf{0.1123} 	\\ \cline{1-4}
\end{tabular}
\vspace{-2mm}
\end{table}

Table~\ref{table:resultsBySessionsType} presents the performance of LTR-SP and other approaches on each of the four session's types. LTR-SP is the best performing approach across all session's types under nDCG@10 and nERR@10. It also achieves the best scores under the MAP metric for all types except for factual specific sessions. This is likely to be caused by the fact that LTR-SP uses category B of ClueWeb12 for TREC 2013 whereas the best TREC system is taking advantage of the full collection. \par

All approaches seem to be excelling at exploratory sessions, which is expected due to the task nature. The task of session search focuses on utilizing a session's data to improve performance for the current query. In exploratory search, a user's goal is ill-defined and her search product is intellectual. Therefore, they refine and proceed in their session based on the results that are shown to them and the interaction they might have made with such results. All approaches are taking advantage of these interaction data and, therefore, these types of sessions seem to benefit the most. \par

For all approaches, sessions with well-defined information needs appear to benefit the least particularly for intellectual tasks. One possible explanation is that the additional information that is used by such approaches may cause a drift from the actual need. The good performance on amorphous sessions comes at a risk of drifting for sessions with specific goals. This is, perhaps, one advantage of LTR-SP, which takes a risk-averse approach. First, social expansion terms are extracted in a way that ensure their relevance to both the session and the session's social position to avoid including extraneous terms. Special subsets of these terms are extracted based on their appearance in clicked documents' titles or snippets. These subsets were shown to produce effective features, see table~\ref{table:GiniAnalysis}. Second, LTR-SP uses a relatedness classifier that is more fine-grained in identifying related sessions than the LDA-based approach. Topical classification as in LTR-LDA performs closely to LTR-SP on amorphous sessions because of their exploratory nature but less so when the information need is well-defined. \par
\section{Conclusion}
\label{section:conclusion}
This paper introduces a novel framework to personalize web search sessions under the framework of learning to rank. Existing approaches have either relied on features extracted from the test session only or the query logs. Our framework uses pre-computed user models and maps each test session to its most relevant user models. These models are used to estimate novel learning features for the learning to rank model. We have shown how such features provide a valuable new source of features in session search. Our experiments on four test collections from the TREC session track demonstrate that the proposed approach is statistically superior to current session search approaches. Unlike previous work, the improvement is consistent on all four collections and performance is stable across various session's types. Sessions' interaction data represent rich contextual information for search engines to use in order to personalize results at the session level. We have shown that incorporating them into various session search systems helps improve the relevance for test queries. Our user models are transparent and could be easily explained to users using their natural labels, i.e. social position. \par

  \bibliographystyle{ACM-Reference-Format}
  \bibliography{bibList}

\end{document}